\documentclass[aps,prx,floatfix,twocolumn,notitlepage,superscriptaddress,10pt]{revtex4-2}
\usepackage{xcolor}
\usepackage{grffile}
\usepackage{amsmath, amsthm, amssymb} %, bbold}
\usepackage[normalem]{ulem}
\usepackage{cancel}
\usepackage{bm}
\usepackage{microtype}
\usepackage{hyperref}
\usepackage{setspace}
\usepackage{grffile}
\hypersetup{colorlinks,linkcolor=blue,urlcolor=blue,citecolor=blue}
\usepackage{amsmath}    % need for subequations
\usepackage{amssymb}
\usepackage{graphicx}   % need for figures
\usepackage{verbatim}   % useful for program listings
\usepackage{color}      % use if color is used in text
\usepackage{hyperref}   % use for hypertext links, including those to external documents and URLs
\usepackage[normalem]{ulem}
\usepackage{natbib}
\usepackage{enumitem} 
\usepackage{dsfont} 
\usepackage{physics} % for hbar and other  physics related notations	
\usepackage{lipsum}	

		\newcommand{\rmd}{\,{\mathrm d}}
		\newcommand{\be}{\begin{equation}}
		\newcommand{\ee}{\end{equation}}

		\newcommand{\Id}{\mathbb{I}}
		
\begin{document}

% AUTHORS ----------------------------------------------------
\title{Spectral Properties Versus Magic Generation in $T$-doped Random Clifford Circuits}
    \author{Dominik~Szombathy}
	   \email{szombathy.dominik@edu.bme.hu}
	   \affiliation{Department of Theoretical Physics,  Institute of Physics, Budapest University of Technology and Economics,  M\H{u}egyetem rkp. 3., H-1111 Budapest, Hungary}
	   \affiliation{Nokia Bell Labs, Nokia Solutions and Networks Kft, 1083 Budapest, B\'okay J\'anos u. 36-42, Hungary}
    \author{Angelo~Valli}
        \affiliation{Department of Theoretical Physics,  Institute of Physics, Budapest University of Technology and Economics,  M\H{u}egyetem rkp. 3., H-1111 Budapest, Hungary }
        \affiliation{MTA-BME Quantum Dynamics and Correlations Research Group, Budapest University of Technology and Economics,  M\H{u}egyetem rkp. 3., H-1111 Budapest, Hungary}
    \author{C\u at\u alin Pa\c scu Moca}
       \affiliation{MTA-BME Quantum Dynamics and Correlations Research Group, Budapest University of Technology and Economics,  M\H{u}egyetem rkp. 3., H-1111 Budapest, Hungary}
        \affiliation{Department of Physics, University of Oradea,  410087, Oradea, Romania}
    \author{J\'anos~Asb\'oth}
        \affiliation{Department of Theoretical Physics,  Institute of Physics, Budapest University of Technology and Economics,  M\H{u}egyetem rkp. 3., H-1111 Budapest, Hungary }
    \author{L\'or\'ant~Farkas}
        \affiliation{Nokia Bell Labs, Nokia Solutions and Networks Kft, 1083 Budapest, B\'okay J\'anos u. 36-42, Hungary}
    \author{Tibor~Rakovszky}
        \affiliation{Department of Theoretical Physics,  Institute of Physics, Budapest University of Technology and Economics,  M\H{u}egyetem rkp. 3., H-1111 Budapest, Hungary }
        \affiliation{HUN-REN-BME Quantum Error Correcting Codes and Non-equilibrium Phases Research Group,
Budapest University of Technology and Economics,
M\H{u}egyetem rkp. 3., H-1111 Budapest, Hungary}
    \author{Gergely~Zar\'and}
          \email{zarand.gergely.attila@ttk.bme.hu}
        \affiliation{Department of Theoretical Physics,  Institute of Physics, Budapest University of Technology and Economics,  M\H{u}egyetem rkp. 3., H-1111 Budapest, Hungary }
        \affiliation{MTA-BME Quantum Dynamics and Correlations Research Group, Budapest University of Technology and Economics,  M\H{u}egyetem rkp. 3., H-1111 Budapest, Hungary}

% TIME -----------------------------------------------------
\date{\today}
% ABSTRACT -------------------------------------------------
\begin{abstract}
We study the emergence of complexity in deep random $N$-qubit $T$-gate doped Clifford circuits, 
as reflected in their spectral properties and in magic generation, 
characterized by the stabilizer R\'enyi entropy distribution and the non-stabilizing power of the circuit. 
For pure (undoped) Clifford circuits, a unique periodic orbit structure in the space of Pauli strings 
implies peculiar spectral correlations and level statistics with large degeneracies. 
$T$-gate doping induces an exponentially fast transition to chaotic behavior, described by random matrix theory. 
We compare these complexity indicators with magic generation properties of the Clifford+$T$ ensemble, 
and determine the distribution of magic, as well as the average non-stabilizing power of the quantum circuit ensemble.
In the dilute limit, $N_T \ll N$, magic generation is governed by single-qubit behavior. 
Magic is generated in approximate quanta, 
increases approximately linearly with the number of $T$-gates, $N_T$, 
and displays a discrete distribution for small $N_T$. 
At $N_T\approx N$, the distribution becomes quasi-continuous, and  for $N_T\gg N$ it 
converges to that of Haar-random unitaries, and averages to a finite magic density, 
$m_2$, $\lim_{N\to\infty} \langle m_2 \rangle_\text{Haar} =  1$. 
This is in contrast to the spectral transition, where ${\cal O} (1)$ $T$-gates suffice to remove spectral degeneracies
and to induce a transition to chaotic behavior in the thermodynamic limit. 
Magic is therefore a more sensitive indicator of complexity. 
\end{abstract}

\maketitle

\section{Introduction}

Quantum physics has opened up a complexity frontier: generic quantum dynamics can easily turn a product state 
of many particles or qubits into a state that is practically impossible to represent, let alone study, 
on a classical computer~\cite{jaeger2007classical}. 
A possible source of complexity is quantum entanglement~\cite{chitambar2019quantum} generated between the constituents; 
without this, quantum dynamics is efficiently simulable classically~\cite{jozsa2003role,vidal2003efficient}. 
However, entanglement in itself does not fully characterize this frontier, 
as there exist dynamics that create highly entangled states and yet can be efficiently simulated classically. 
A chief example is that of Clifford circuits, which have an efficient representation in terms of the so-called stabilizer formalism~\cite{Gottesman1998, aaronson2004improved}, 
%while they generate
despite generating entanglement~\cite{nahum2017quantum,kim2013ballistic}. 
%Clifford circuits take any $N$-qubit basis state to an $N$-qubit stabilizer state, 
%which can be highly entangled yet efficiently represented using its $N$ Pauli stabilizers. 
%Because of their efficient simulability, 
For these properties, Clifford circuits are often used as easily simulable toy models of 
quantum many-body dynamics, where various quantities of interest can be 
studied numerically~\cite{chandran2015semiclassical,nahum2017quantum,znidaric2020entanglement}.

While Clifford circuits exhibit certain properties similar to generic quantum dynamical systems, 
their efficient simulability comes with some highly non-generic qualities~\cite{keyserlingk2018operator}. 
A possible tool to assess the lack of complexity of a circuit is provided by \emph{spectral correlations}.
As we demonstrate in this work, the spectral properties of Clifford circuits are very peculiar: the spectrum
does not display level repulsion, characteristic to chaotic systems or Haar-random circuits, 
although it is not Poissonian either.

Clifford circuits can also be regarded as lacking complexity, in connection with the notion 
of \textit{non-stabilizerness}. Universal quantum computation can be achieved only relying on 
non-Clifford operations.  States that cannot be generated by Clifford circuits from computational basis states 
exhibit a quantum resource referred to as non-stabilizerness, or "magic" ~\cite{veitch2014resource}. 
Among  non-stabilizerness measures that have been proposed~\cite{veitch2014resource,Howard2017,Heinrich2019,Wang2019,Wang2020,Beverland2020,Liu2022},  
the recently introduced \textit{stabilizer R\'enyi entropy} (SRE)~\cite{Leone2022} 
raised significant attention \cite{Oliviero2022,Rattacaso2023,Haug2023,Lami2023,Bejan2024,Haug2024,Tarabunga2024,Niroula2024,Turkeshi2024,gidney2024magic,Fux2410.09001,Turkeshi2407.03929}.
The simplicity or complexity of a \emph{circuit} can be assessed by examining 
the magic of the states it produces from basis states, 
and quantified in terms of SRE by the \emph{non-stabilizing power}~\cite{Leone2022}.

In this regard, it has been known for a long time that extended ``Clifford+$T$-gate sets" 
realize a platform for universal quantum computation,
and can therefore be used to approximate arbitrary unitary operations~\cite{Kitaev1997}, 
while constantly looking for optimized strategies to implement $T$ gates~\cite{Litinski2019,litinski2019magic,gidney2024magic}. 
%
%The realization of $T$-gates is thus an important task for fault-tolerant quantum computation, 
%where this is often done using magic states and gate teleportation, 
%and often consumes much of the resources of a computation~\cite{Litinski2019,litinski2019magic,gidney2024magic}. }
%
%The transition from the highly nongeneric Clifford circuits to generic (chaotic) circuits can be realized by injecting so-called $T$-gates. 
Interestingly, while adding a small number of $T$-gates does not spoil classical simulability~\cite{bravyi2016improved}, 
it is already sufficient to make a quantum circuit mimic a generic unitary ensemble~\cite{Haferkamp2022}, 
and even a single $T$-gate in a Clifford circuit can drive a transition 
to universal entanglement spectrum statistics~\cite{Zhou2020,trueQuantum6}, 
{although} some universal features are only recovered for an extensive, i.e., $O(N)$, number of $T$ gates~\cite{trueQuantum6}. 

In this work, we address the question, how introducing $T$-gates into Clifford circuits induces complexity by the 
above measures. 
%In this work, we perform a detailed study of the spectral properties of random Clifford circuits 
%and how they evolve when the circuit is doped with $T$-gates. 
%In this work, we characterize various properties of random Clifford circuits. 
%In this work, we study how various properties of Clifford circuits behave under $T$-doping. 
Instead of limiting ourselves to specific examples, 
we consider an ensemble of random Clifford circuits and study its statistical properties  
to uncover typical behavior, while focusing on two different but complementary aspects: 
(i) the structure and evolution of the peculiar Clifford spectrum, 
%characterized by high eigenvalue degeneracy, 
and (ii) the way magic is generated by injected $T$-gates. 
We find that non-stabilizerness and spectral correlations of random Clifford circuits reveal 
increasing complexity upon ``doping"  with $T$-gates.

First, to uncover spectral properties of pure Clifford circuits with no $T$-gates, we relate spectral correlations to 
the distribution of periodic orbits in the space of Pauli strings. 
Dynamics there decomposes into periodic orbits, giving rise to a peculiar ``Clifford spectrum'', 
which exhibits large degeneracies, while it is clearly distinct from a Poissonian spectrum usually 
associated with integrable systems.
Introducing $T$-gates disrupts these orbits, the peculiar Clifford spectrum gradually disappears,
and evolves into a generic chaotic behavior described by random-matrix theory (RMT), 
as signalled by various indicators. 
Our results suggest that an $\mathcal{O}(1)$ number of $T$-gates suffices to make the average spectral properties RMT-like, resembling the scaling observed for other measures of chaos~\cite{Haferkamp2022,Zhou2020}. 
%This must be contrasted with certain 
%features of \emph{stabilizerness}, which are known to \emph{require a finite density} of $T$ 
%gates to disappear \cite{Hamma paper}. 
This behavior is somewhat analogous to the property of perturbed classical  integrable systems, 
where -- according to the KAM theorem~\cite{poschel2009lecture} -- a small 
nonintegrable perturbation cannot entirely destroy periodic orbits in the entire phase space. 
  
We compare this behavior with the recently-introduced circuit complexity indicator,   
the non-stabilizing power of an operator~\cite{Leone2022}, 
which characterizes the magic generation power of the circuit~\cite{vairogs2024extracting,zhang2024quantum}
by using the notion of the 
\emph{stabilizer Renyi entropy} (SRE)~\cite{Leone2022,wang2023stabilizer,haug2023stabilizer}, 
a practical quantifier of magic. 
%
%However, instead of focusing on the magic contained in a state, we concentrate on characterizing the 
%magic (generation) properties of specific
% \textit{ensembles of quantum operations}~\cite{vairogs2024extracting, zhang2024quantum}.
By a statistical sampling of deep Clifford+$T$ circuits, we obtain the corresponding SRE (magic) distribution  
and investigate its evolution as a function of the number of injected $T$-gates, $N_T$. 
Averaging the magic distribution {over the initial stabilizer states} also allows us to compute 
{the average magic of the final state, which is equivalent to}
the average non-stabilizing power of the circuit ensemble, $\langle {\cal M}_2\rangle$.
We find that, in the dilute limit, $N_T \ll N$, with $N$  the number of qubits, magic is 
typically generated in discrete steps (quanta), and the average non-stabilizing power increases linearly 
with the number of $T$-gates in large circuits. 
Correspondingly, magic distribution is discrete for small values of $N_T$, while a continuous  SRE distribution emerges
for an extensive number of injected $T$-gates, $N_T\sim N$ and, 
eventually, it converges to that of the Haar random unitary ensemble. 
Our results agree with those of Ref.~\cite{haug2024Probing}, 
where a phase transition has been predicted above a critical density of $T$-gates, $n_T^{*}\equiv (N_T/N)^* \approx 2.41$. 
Below this value, we confirm that the average magic density 
%$\langle M_2/N\rangle = \langle {\cal M}_2/N\rangle$ 
is simply proportional to the density of $T$-gates, 
while it
%generated by the circuit 
approaches its maximal value  %$\langle M_2/N\rangle\to 1$ 
beyond this concentration. 
%, while below this critical density magic density is just proportional to $n_T$. 
These findings are in line with the observation that certain 
features of \emph{stabilizerness} require \emph{a finite density} of $T$ gates to disappear~\cite{trueQuantum6}.

The manuscript is organized as follows. 
In Sec.~\ref{sec:circuits} we discuss the anatomy of the random circuits considered in this work. 
Section~\ref{sec:spectral_properties} is devoted to the analysis of the spectral properties.  
In Sec.~\ref{sec:spectrum_Clifford} we recall the relevant notions of Clifford algebra 
to derive the peculiar structure of the periodic orbits and its relation to the eigenvalue spectrum, 
whereas in Sec.~\ref{sec:spectrum_Clifford+$T$} we discuss the effects of injecting $T$-gates 
on the the eigenvalue correlation function and the level spacing statistics. 
Sec.~\ref{sec:magic_generation} focuses on magic. 
We start from the paradigmatic case of a single qubit in Sec.~\ref{sec:magic_one-qubit}, 
which holds the key to understand the magic generation mechanism Clifford+$T$ circuits, 
as discussed in Sec.~\ref{sec:magic_Clifford+$T$}. 
Finally, Sec.~\ref{sec:discussion} contains our conclusion and an outlook.

\section{T-doped random Clifford circuits} \label{sec:circuits}

\begin{figure}[h!]
    \centering
    \includegraphics[width=0.99\linewidth]{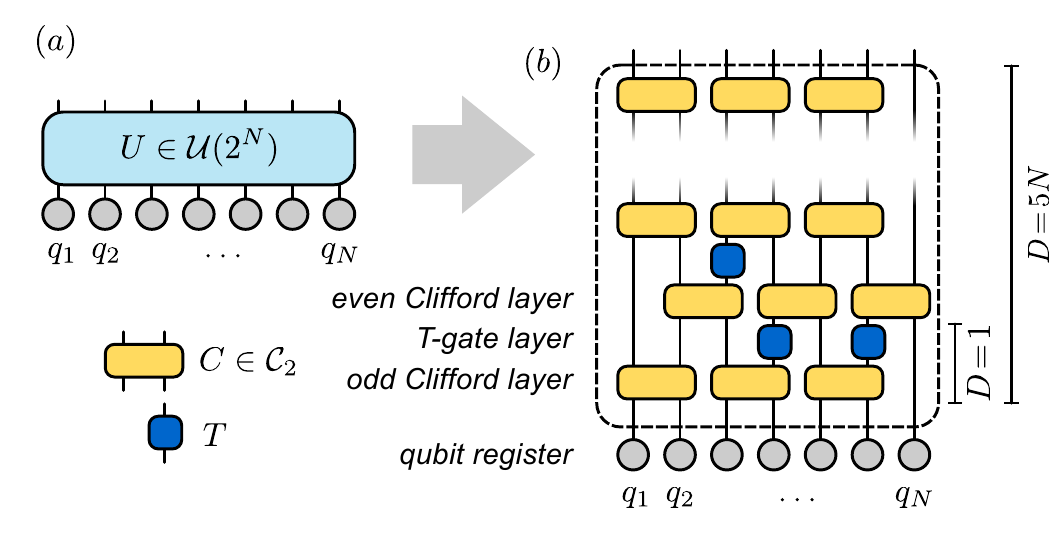}
    \caption{(a) Unitary operator $U \in \mathcal{U}(2^N)$ operating on a $N$-qubit register $\{q_1, q_2, \ldots, q_N\}$. 
    (b) Schematic construction of a brick-wall random Clifford$+T$ gates circuit. 
    Each layer includes $2$-qubit Clifford gates $C \in \mathcal{C}_2$ acting on neighboring qubits 
    and ({for doped circuits}) a layer of randomly \textit{injected}  $T$-gates. 
    Typical circuits considered in the following {contain} $D=5N$ such layers (see text for details). }
    \label{fig:BWschematics}
\end{figure}

%\tibor{Changed the title of this section.}

Let us start by characterizing the class of random circuits under consideration, 
%which is 
schematically represented in Fig.~\ref{fig:BWschematics}. 
Given an array of $N$ qubits, we build circuits with a \textit{brick-wall} structure, 
i.e., composed of odd and even layers of $2$-qubit Clifford gates $C \in \mathcal{C}_2$. 
Each gate is chosen randomly with uniform probability from all possible $|\mathcal{C}_2|=11520$ 2-qubit Clifford gates. 
%\tibor{which} consists of $|\mathcal{C}_2|=11520$ elements. 
The depth $D$ of a circuit is given by the number of layers. 
Additionally, after each Clifford layer, we include another layer in which we inject $T$-gates ($\pi/4$ phase gates). 
For a given total number $N_T$ of  $T$-gates in the circuit, 
both the layer index $l$ and the qubit index $i$ on which the $T$-gate acts are chosen randomly, 
with a constraint preventing $T$-gates to have the same $(l,i)$ pair, 
since the action of two consecutive  $T$-gates corresponds to a phase gate $T^2=S$, 
which belongs to the Clifford group. 

%\tibor{I replaced ``brickwall'' with ``brickwork'' everywhere}

We note that the particular choice of brick-wall architecture is largely unimportant for our discussion below. 
We will be interested in the limit of deep circuits, $D \gg N$, 
and in this limit the random brick-wall Clifford-circuit ensemble becomes indistinguishable 
from the uniform distribution over \emph{all} $N$-qubit Cliford unitaries. 
Therefore, the results that we obtain for (deep) brick-wall construction are representative of the Clifford group $\mathcal{C}_N$. 
The necessary depth $D$ needed to achieve this limit is related to entanglement spreading. 
In particular, we expect a linear growth of entanglement, 
characterized by an ``entanglement velocity''~\cite{mezei2017entanglement,von2018operator}, 
which sets the number of layers needed to entangle all qubits; 
this required depth is therefore expected to scale linearly with the system size $N$. 
We verified numerically that the spectral properties of the Clifford circuits converge with respect to the circuit depth. 

The choice of the brick-wall structure provides a number of advantages.
In particular, it is computationally efficient to 
(i) build  and apply the \textit{tableau} determining the action of each $2$-qubit Clifford gates
on strings, and  (ii) to sample a collection of $2$-qubit gates rather than a single $N$-qubit gate, 
since such operation scales with $\mathcal{O}(N^2)$ for state-of-the-art algorithms~\cite{Bravyi2021,VanDenBerg2021}.

\section{Spectral properties of doped Clifford circuits} \label{sec:spectral_properties}

Our goal is to characterize the spectral properties 
of random $T$-doped brick-wall Clifford circuits (henceforth Clifford circuits).
Before turning to the $T$-doped case, let us investigate the 
rather peculiar spectral properties of undoped Clifford circuits. 

\subsection{Spectral properties of pure Clifford circuits} \label{sec:spectrum_Clifford}

%\subsubsection{Periodic Orbits} \label{sec:orbits}
%It is useful to recall some definitions. 
An $N$-qubit Clifford operator $C$ acts on the $2^N$dimensional  
%can be represented as a $2^N\times2^N$ matrix on the 
Hilbert space of the $N$ qubits. 
%a tensor product of the computational basis states $\ket{0}$ and $\ket{1}$ on each qubit. 
A Clifford operator takes Pauli strings $S$ to Pauli strings via conjugation (up to a global phase), $S\to C\, S\, C^\dagger$.
More formally, the Clifford group $\mathcal{C}_N$ acting on a $N$-qubit register is defined as \\
{
\begin{equation} 
 \mathcal{C}_N = \left\{ C \in \mathcal{U}(2^N) \mid C \tilde{\mathcal{P}}_N  C^{\dagger} = \tilde{\mathcal{P}}_N \right\} /\,\mathcal{U}(1)
\end{equation}
where $\tilde{\mathcal{P}}_N$ denotes the set of {\textit{signed}} $N$-qubit Pauli strings, 
}
whose elements are generated by the tensor product of single-qubit \textit{signed} Pauli matrices,
and {$\mathcal{P}_N = \tilde{\mathcal{P}}_N/\langle\pm 1\rangle$} is defined as the 
set of \textit{unsigned} Pauli strings. 
The operators  $C$ are elements of {the unitary group of $N$-qubit operations $\mathcal{U}(2^N)$}, and are 
defined up to an {overall $\mathcal{U}(1)$ phase }. 
%Then, $\mathcal{C}_N$ is the quotient group by the subgroup of scalar unitary matrices $\mathcal{U}(1)$. 
%general definition from: \url{https://docs.quantum.ibm.com/api/qiskit/qiskit.quantum_info.Clifford}]
%
%

The action of a Clifford operator $C \in \mathcal{C}_N$ on an {unsigned} Pauli string $S_i \in \mathcal{P}_N$
is given by
 \begin{equation}
     \pm S_{i+1} = C S_i C^\dagger,
 \end{equation}
where {$S_{i+1} \in \mathcal{P}_N$}, 
%and the indices $j=i,i+1,\dots$ label the series of Pauli strings obtained by the action of a Clifford operator. 
{and the indices $i$ and $i+1, \dots$ label unsigned Pauli strings obtained after consecutive actions of the Clifford gate $C$.}
Since the dimension $|\mathcal{P}_N|$ of the {set of unsigned Pauli strings} is finite, 
the repeated action of the \textit{same} element $C$ results in periodic orbits of period $L$, defined by the condition 
 \begin{equation}
 %    \underbrace{C \cdots C}_{L} S_0 \underbrace{C^\dagger \cdots C^\dagger}_{L} = 
     C^L S_0 (C^\dagger)^L = \mathbb{C}^L S_0 = \tau S_0\;,
 \end{equation}
where we introduced the notation $\mathbb{C} \,O \equiv C O C^\dagger$ with $O$ an operator acting on the 
%Hilbert space of 
qubits.  
On an orbit of length $L$, the initial Pauli string $S_0$ is recovered after $L$ steps – up to an overall sign $\tau = \pm 1$.
For any given Clifford operator $C$, the number of orbits $N_C (L, \tau)$ with period $L$ and parity $\tau$ 
fulfills the sum rule 
\begin{equation}
 \sum_{L, \tau} \tilde{N}_C (L, \tau) = \sum_{L, \tau}  L N_C (L, \tau) = |\mathcal{P}_N| = 4^N,
\end{equation}
where 
%is the probability 
$\tilde{N}_C (L, \tau)$ denotes the total number of those strings 
that are on some orbit of length $L$ and parity $\tau$.
%
%We can further define the ensemble averaged probability of having an orbit with period $L$ and parity $\tau$ as 
% \begin{equation}
%    P(L, \tau) = \left\langle P_C(L,\tau) \right\rangle_C, 
% \end{equation}
%where $\langle \cdots \rangle_C$ denotes the average over Clifford operators.  
{We can further define the probability for a Clifford operator $C$ to allow for an orbit of length $L$ and parity $\tau$ as 
$P_C(L,\tau) = LN_C(L,\tau)/4^N$ and the corresponding ensemble-averaged probability as 
 \begin{equation}
    P(L, \tau) = \left\langle P_C(L,\tau) \right\rangle_C, 
 \end{equation}
where $\langle \cdots \rangle_C$ denotes the average over Clifford operators. 
}
 \begin{figure}[t!]
 \begin{center}
    \includegraphics[width=0.95\columnwidth]{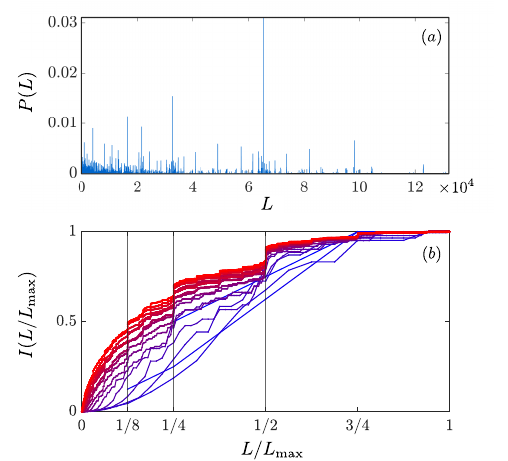}
    \caption{(a) Parity-integrated probability $P(L)$ for $N = 16$. 
    (b) Integrated probabilities $I(L/L_{\mathrm{max}})$ for $N = \{2, 3, \dots, 16\}$ (from blue to red). 
    The vertical black lines denote representative rational values $L/L_{\mathrm{max}}=p/q$.}
    \label{fig:orbits}
 \end{center}
\end{figure}
In Fig.~\ref{fig:orbits}~(a) we show the probability of Clifford orbits $P(L)=\sum_\tau P(L,\tau)$ for $N=16$ qubits. 
%For $N=2$ we show the \textit{exact} probability $P(L)$ and the parity-resolved probabilities $P(L,\tau)$, 
%obtained by calculating all orbits for all Clifford gates $C \in \mathcal{C}_2$. 
Statistical sampling has been performed over brick-wall Clifford circuits of fixed depth $D$ 
as schematically shown in Fig.~\ref{fig:BWschematics} (without $T$-gates at this stage).  
As stated before, for deep-enough circuits 
this is {equivalent to sampling} 
generic $N$-qubit Clifford {unitaries} $C \in \mathcal{C}_N$. 
The distribution of probabilities $P(L)$ is  denser for short orbits while it is sparser for large orbits, 
and displays some nontrivial properties.  
Orbits of length $L$ come with a large degeneracy. While there are $4^N$ string configurations, 
the largest orbit is found to scale only as $L_{\mathrm{max}} = 2^{N+1}$. 
Moreover, the number of \emph{allowed} {orbit lengths, $L$,} %circuits  
is exponentially smaller than $L_{\mathrm{max}}$. 
Rather surprisingly, the largest probability is found for $L = L_{\mathrm{max}}/2$, and 
atypically-high probabilities occur at rational multiples of $L_{\mathrm{max}}$. 
Note that as the system size increases, it becomes statistically less likely to sample orbits of length $L\sim \mathcal{O}(1)$. 

The peculiar distribution of the orbits becomes more apparent by plotting the integrated probability 
$I(L/L_{\mathrm{max}})$ for different $N$, see Fig.~\ref{fig:orbits}(b).
$I(L/L_{\mathrm{max}})$ displays steep jumps at rational multiples of $L/L_{\mathrm{max}} = p/q$. 
%with $p, q \in \mathbb{Z}_{+}$ and $p < q$. 
%The most evident are found at inverse powers of $2$, i.e., for $p=1$ and $q=2^k$ with $k \in \mathbb{Z}_{+}$. 
The most pronounced jumps occur at $L/L_{\mathrm{max}} = \{1/2, 1/4, 1/8, 1/16, \ldots \}$ but
they also appear at fractions $\{3/8, 5/16, 7/16, \ldots \}$, see Fig.~\ref{fig:orbits}(b). 
%There does not appear to be a clear relation between the ratio and the height of the corresponding jump. 
%Moreover, we find that 
With increasing $N$ the jumps become sharper 
but their height appears to scale to zero in the thermodynamic limit $N \to \infty$, 
suggesting that $I(L/L_{\mathrm{max}})$ may converge to a continuous curve.

%\subsubsection{Structure of the Eigenvalues} \label{sec:eigenvalues}

The structure of the periodic orbits allows us to establish a connection 
between the eigenvalue spectrum of Clifford operators $C$ and $\mathbb{C}$. 
For a given orbit, $\{S_0, \mathbb C S_0,\ldots, {\mathbb C}^{L-1} S_0\}$, we can construct the set of operators 
\begin{equation}
    V_m = \sum_{k = 0}^{L - 1} e^{i \frac{2 \pi}{L}\, k\, m}\, \mathbb{C}^k S_0,
\end{equation}
with %the parity-dependent coefficient
$m \in \{0,1,\dots, L-1\} $ for even parity orbits, $\tau=1$, 
and $m \in \{\frac{1}{2}, \dots L- \frac{1}{2}\}$ for $\tau=-1$. 
%Acting with $\mathbb{C}$ on $V_m$ yields
%\begin{equation}
%    \mathbb{C} V_m =  \sum_{k^{\prime} = 1}^{L} e^{i \frac{2 \pi}{L}\, k^{\prime}\, m}\, \mathbb{C}^{k^{\prime}} S_0,
%\end{equation}
Clearly, the $V_m$'s are eigenvectors of $\mathbb{C}$,  
\begin{equation}
    \mathbb{C}V_m = e^{i \Theta_m}V_m \quad \text{with} \quad  \Theta_m = \frac{2 \pi}{L} m. 
\end{equation}
%with an eigenphase 
%\begin{equation}
%    \Theta_m = \frac{2 \pi }{L} m.
%\end{equation}
%
The eigenvalues of $\mathbb{C}$ are thus related to the periodic orbits. 
There are altogether $4^N$ such eigenvalues, corresponding to the dimension of the space of unsigned Pauli strings. 

\begin{figure}[tp!]
    \includegraphics[width=0.95\columnwidth]{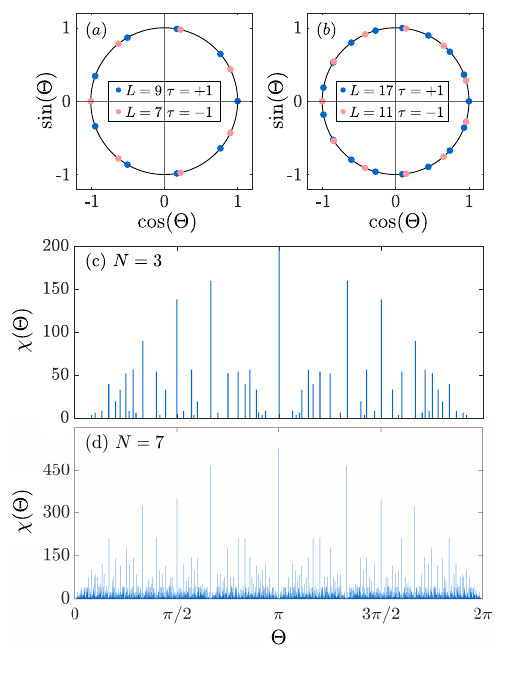}
    \caption{(a,b) Representative eigenvalue distribution of the Clifford operator $\mathbb C$ on the unit circle, 
    corresponding to orbits of different length $L$ and parity $\tau$.
    (c,d) Phase correlation function for $N = 3$ and $N = 7$ random Clifford circuits. 
    While collecting the data, we used box sizes $\Delta \Theta = 2\pi / 16000$ and  $\Delta \Theta = 2\pi / 256000$. 
     Note the common structure with the highest peaks located at rational multiples of $\pi$ (see text for a discussion). 
    Histograms were obtained by sampling up to $N_s=2^{16}$ circuits. }
    \label{fig:chiClifford}
\end{figure}

On the other hand, these eigenvalues (and the phases $\Theta_m$) 
are also closely related to the $2^N$ eigenvalues (spectrum) of the Clifford operator $C$. 
In particular, let us denote the eigenvectors and eigenvalues of $C$ by 
\begin{equation}
    C\ket{\phi_i} %= \lambda_i \ket{\phi_i} 
    = e^{i\vartheta_i}\ket{\phi_i}. 
\end{equation}
Then  $\ket{\phi_i}\bra{\phi_j}$ is an eigenstate of the operator $\mathbb{C}$ with eigenvalue $\Theta = \vartheta_i-\vartheta_j$, 
\begin{equation}
    \mathbb{C}\ket{\phi_i}\bra{\phi_j} = C \ket{\phi_i}\bra{\phi_j} C^\dagger = e^{i(\vartheta_i - \vartheta_j)}\ket{\phi_i}\bra{\phi_j}.
\end{equation}
The structure of the periodic orbits is naturally reflected in the distribution of the corresponding phases $\theta_j$ on the unit circle.
The eigenvalues corresponding to representative periodic orbits of length $L$ and parity $\tau$ are shown in Figs.~\ref{fig:chiClifford}(a,b). 

Notice that the relation above, relating  the spectrum of a unitary operator $U$ and that of the corresponding operator 
$\mathbb U$ is general. The spectrum of $\mathbb U$ thus captures spectral correlations in the spectrum of $U$. 
To investigate the properties of the phase distribution, 
we define the phase correlation function for an ensemble of unitary operators $U \in \mathcal{U}(2^N)$ as
%\begin{align} \label{eq:chi}
%   \chi(\vartheta, \vartheta^\prime) &= \langle \varrho_U(\vartheta) \varrho_U(\vartheta^\prime)\rangle_U^{\rm{conn}} \nonumber \\ 
%                                                      &= \frac{1}{2\pi \, d(d-1)}\sum_{n \neq m} 
%                                                           \langle \delta_{2\pi} \big( \vartheta - \vartheta_n(U)              \big) 
%                                                                       \delta_{2\pi} \big( \vartheta^\prime - \vartheta_m(U) \big) \rangle_U ,
%\end{align}
\begin{widetext} 
\begin{equation}
\label{eq:chi}
   \chi(\vartheta, \vartheta^\prime) 
   %this is not true
   %= \langle \varrho_U(\vartheta) \varrho_U(\vartheta^\prime)\rangle_U
                                                      = \frac{1}{d(d-1)} \sum_{i \neq j=1}^{d}
                                                          \langle \delta_{2\pi} \big( \vartheta - \vartheta_i(U)              \big) 
                                                                      \delta_{2\pi} \big( \vartheta^\prime - \vartheta_j(U) \big) \rangle_U ,
\end{equation}
\end{widetext}
where $\langle \ldots \rangle_U$ indicates the statistical average over $U$, and 
$d=2^N$ is the dimension of the Hilbert space.  
Here $\delta_{2\pi}(\ldots)$ denotes the $2\pi$-periodic Dirac $\delta$ function. 
Notice that we have removed the contribution of the terms with $i=j$, and normalized $\chi(\vartheta,\vartheta')$ as 
%\be
$
\int_0^{2\pi} \rmd \vartheta \int_0^{2\pi} \rmd \vartheta' \;    \chi(\vartheta, \vartheta^\prime) = 1\;.
$
%\ee

%For a given – not necessarily Clifford – circuit, the phase density is defined as
%\begin{equation}
%   \varrho_U ( \vartheta ) = \frac{1}{d} \sum_{j=1}^{d} \delta_{2\pi}(\vartheta - \vartheta_j(U)).
%\end{equation}
%with $\delta_{2\pi}(\ldots)$ denoting the $2\pi$-periodic Dirac $\delta$ function. 

For Haar-random unitaries, $\chi(\vartheta,\, \vartheta^\prime)$ depends only on the phase difference. 
To compare spectral correlations in a random Clifford circuit with those in a Haar-random circuit, we therefore introduce 
the average spectral correlation function as 
\be
\chi(\Theta)\equiv \int_0^{2\pi} \rmd \vartheta' \;    \chi(\Theta +\vartheta^\prime, \vartheta^\prime)\;.
\ee
%Averaging over $U$, implies that, If the global phase of the ensemble of unitary operators $U$ is random 
%-- which is the case of random Clifford circuits or random unitary circuits -- 
%the correlation function $\chi$ depends only on the phase difference
%\begin{equation}
%$
%    \chi(\vartheta,\, \vartheta^\prime) = \chi(\vartheta - \vartheta^\prime), 
     %\rightarrow \chi(\Theta)
%\end{equation}
%and is normalized as
%\begin{equation}
 %$\int_{0}^{2\pi} d\vartheta \, \chi(\Theta) = 1$. %\sum_{j=1}^{N_b} \Delta b\, \chi(\Delta\vartheta_j) = 1, 
%\end{equation}
%where $N_b$ denotes the number of bins of size $\Delta b = 2\pi/N_b$, and $\Delta\vartheta_j$ is the value at the center of the $j$-th bin. 
%Operatively, the eigenvalues are calculated by diagonalizing the operator $U$ in its Hilbert space representation,  
%and the corresponding phase differences are binned in a histogram. 

Figs.~\ref{fig:chiClifford}(c,d) compare the correlation function of undoped Clifford circuits for different system sizes, $N$, 
revealing the peculiar structure and common features of $\chi(\Theta)$. 
The most prominent peaks are observed at rational multiples of $\pi$, including $\{\pi, 2\pi/3, \pi/2, \pi/3, \pi/4, \pi/8, \ldots\}$ 
which are typically surrounded by regions of lower peak density and intensity. 
These peaks get contribution from a large number of orbits. 
For example, all $L=\text{even}$ and $\tau=1$
orbits as well as all $L=\text{odd}$ and $\tau=-1$ orbits
have a correlation peak at $\Theta=\pi$, while all $L=0 \mod 3$ orbits with $\tau=1$ have a peak at $\pm 2\pi/3$.
These peaks, especially the peak at $\Theta =\pi$, are sensitive indicators of periodic orbits.
Although our analysis was not conclusive in this regard, the structure observed is reminiscent of a fractal.

\subsection{T-doped Clifford circuits} \label{sec:spectrum_Clifford+$T$}

Extending the Clifford group with a single non-Clifford gate (e.g., $T$-gate) is sufficient to achieve universal quantum computing~\cite{bravyi2005universal,bravyi2016improved,hinsche2023one}. 
Hence, the natural goal is to understand how the spectral properties of Clifford circuits change 
upon \textit{injecting} $T$-gates in the random circuit (henceforth Clifford+$T$ circuits). 
The expectation is that the circuit properties should converge towards those of 
Haar distributed operators of the circular unitary ensemble (CUE)~\cite{deift2009random,gharibyan2018onset,Forrester2023}. 
In this respect, a relevant problem is to quantify \textit{how many} $T$-gates are necessary for this convergence. 
To this end, in the following we consider two quantities: (i) the phase correlation function defined in Eq.~(\ref{eq:chi})  
and (ii) the level spacing statistics. 

\begin{figure}[tp!]
    \centering
    \includegraphics[width=0.95\columnwidth]{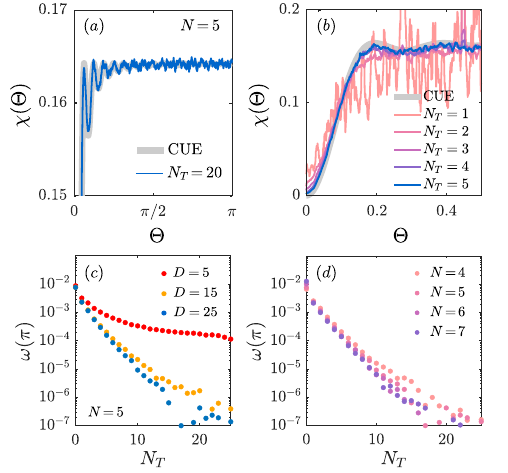}
    \caption{(a) Comparison of a representative correlation function $\chi(\Theta)$
    for a random Clifford circuit of $N=5$ qubits with $N_T=20$ injected $T$-gates and of depth $D=25$ with (blue) 
    against the corresponding analytical result from RMT for the CUE from Eq.~(\ref{chiRMT}) (grey). 
    (b) Evolution of $\chi(\Theta)$ with an increasing number $N_T$ of $T$-gates for a random Clifford circuit of $N=5$ qubits, and 
    (c) suppression of the weight $\omega(\pi)$ of the delta peak at $\Theta=\pi$ with the 
    number of $T$-gates for different values of the circuit depth, $D=5$, $D=15$, and $D=25$. 
    %The grey line denotes the value $\chi_{\mathrm{CUE}}(\pi)$. 
    (d) Suppression of the weight of the $\Theta=\pi$ peak as a function of $N_T$ for different values of $N$ and $D=5N$. }
    \label{fig:vsRMT}
\end{figure}

\subsubsection{Phase correlation function} \label{sec:correlation_function}

The analytic expression for the correlation function of the CUE is known from RMT~\cite{gharibyan2018onset}
\begin{equation} \label{chiRMT} 
     \chi_{\mathrm{CUE}}(\Theta =\vartheta-\vartheta^\prime) = \frac{1}{2 \pi} \frac{d}{d-1} \left(1 - \frac{\sin^2( d \,\Theta /2 ) }{d^2 \sin^2 (\Theta/2)} \right) 
\end{equation}
For a high number of $T$-gates $N_T \gg 1$, we find that the correlation function $\chi(\Theta)$ of a $T$-doped 
Clifford circuits resembles very closely $\chi_{\mathrm{CUE}}(\Theta)$ as shown in Fig.~\ref{fig:vsRMT}(a), 
in particular, the characteristic rapid oscillations towards the edge $\vartheta \sim 0$ are quantitatively reproduced. 

\begin{figure}[tp!]
    \centering
    \includegraphics[width=0.99\columnwidth]{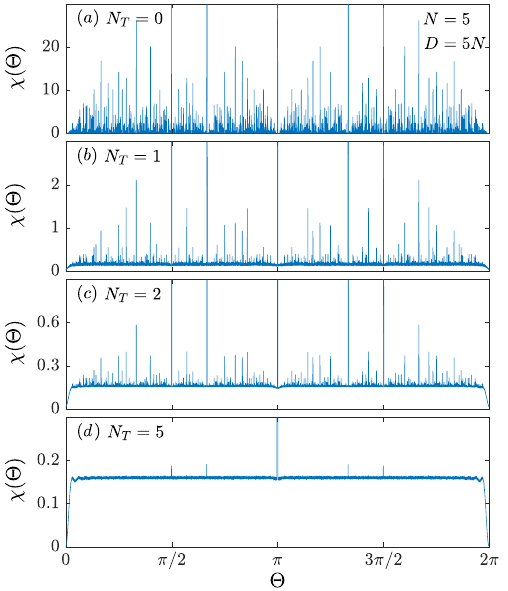}
    \caption{Evolution of the correlation function for $N = 5$ qubits with the number of  $T$-gates $N_T$.
     We used box sizes $\Delta \Theta = 2\pi / 128000$.}
    \label{fig:chiTdoped}
\end{figure}

Fig.~\ref{fig:chiTdoped} displays the evolution of the correlation function with a few injected  $T$-gates. 
The  general structure of $\chi$ is significantly changed already with \textit{a single}  $T$-gate. 
Moreover, the characteristic edge features of $\chi_{\mathrm{CUE}}$ are already recovered 
\textit{qualitatively} with one  $T$-gate and \textit{quantitatively} with as few as three  $T$-gates, 
cfr. also Fig.~\ref{fig:vsRMT}(b).  

In the central region, $\Theta \sim \pi$, the correlation function $\chi$ is dominated by discrete peaks 
reminiscent of the structure observed for pure Clifford circuits. 
These are likely associated with short periodic orbits localized in a portion of the qubit register 
that statistically survive the effect of a few randomly injected  $T$-gates. 
The expectation is that for deep-enough circuits a few $T$-gates should scramble most periodic orbits. 
This is confirmed by the paradigmatic evolution of the feature at $\Theta = \pi$, see Fig.~\ref{fig:vsRMT}(c). 
Notice that already a single $T$-gate suppresses exponentially the correlation peak at $\pi$, i.e., it removes most
 periodic orbits, but an exponentially small fraction of periodic orbits seems to survive even in the large $D$ limit. Also, as a function of 
 $N_T$, data for different $N$-s appear to collapse to a single curve, suggesting that the fraction of removed periodic orbits depends just on $N_T$ for deep circuits. 

\subsubsection{Level spacing statistics} \label{sec:level_spacing}

Another natural measure of how rapidly Clifford+$T$ circuits converge to the CUE 
is given by level spacing statistic ~\cite{carmeli1974statistical,zyczkowski1994random,trueQuantum6}. 
For any unitary operator $U \in \mathcal{U}(2^N)$, we calculate all $d=2^N$ eigenvalues,  
%of the corresponding Hilbert operator, 
sort them (so that $\vartheta_j < \vartheta_k$ for $j<k$), 
and look at the nearest-neighbor %(NN) 
distances $\zeta \equiv \vartheta_{j+1} - \vartheta_{j}$. 
Ensemble averaging over $N_s$ samples yields
\begin{equation}
P(\zeta) = \frac{1}{N_s d}\sum_{i=1}^{N_s} \sum_{j=1}^{d} 
               \delta_{2\pi} \left( \zeta - \left[ \vartheta_{j+1}(U_i) - \vartheta_j(U_i) \right] \right).
% \chi_{\mathrm{NN}}(\vartheta) = \frac {1}{N_s d}\sum_{i=1}^{N_s} \sum_{j=1}^{d} 
 %               \delta_{2\pi} \left( \vartheta - \left[ \vartheta_{j+1}(U_i) - \vartheta_j(U_i) \right] \right),
\end{equation}
%and we denote by $\tilde{\chi}_{NN}(\vartheta)$ the normalized one.
%\dom{\begin{equation}
%\tilde{\chi}_{\mathrm{NN}}(\Delta\vartheta) = \frac{M}{2\pi}\frac{1}{N_S}\frac{1}{d - 1}\sum_{i=1}^{N_s} \sum_{j=1}^{d} 
%                \delta\left( \Delta\vartheta - \left[ \vartheta_{j+1}(U_i) - \vartheta_j(U_i) \right] \right). 
%\end{equation}}
%An approximate for of the distribution is given by
%\begin{equation}
% P_{\mathrm{CUE}}(s) \approx \frac{32}{\pi^2} s^2 \exp \left(-\frac{4}{\pi}s^2\right),
%\end{equation}
%where we introduced the \textit{normalized} level spacing 
%\begin{equation}
% s = \frac{\vartheta_{j+1}-\vartheta_j}{\left\langle \vartheta_{j+1}-\vartheta_{j} \right\rangle}. 
%\end{equation} 
%Assuming we sample enough eigenvalues, we can expect the eigenvalues to be equidistantly distributed on the unit circle
%\begin{equation}
% \left\langle \vartheta_{j+1}-\vartheta_{j} \right\rangle \to \frac{2\pi}{2^N},
%\end{equation}
%thus yielding the normalized distribution
%\begin{equation}
% P_{\mathrm{NN}}(\Delta\vartheta) = \frac{4}{\pi^5} (\Delta\vartheta)^2\, 2^{3N} \exp \left(-\frac{2^{2N}}{\pi^3} (\Delta\vartheta)^2 \right). 
%\end{equation}
%However, for finite $N$ the Wigner-Dyson distribution is approximate, 
%and a better strategy is to sample both random Clifford+$T$ circuits and Haar random unitaries 
%to build the corresponding distributions $\chi_{\mathrm{NN}}$ numerically.  
We sample both random Clifford+$T$ circuits and Haar random unitaries 
to build the corresponding distributions numerically.  
\begin{figure}[tp!]
    \begin{center}
    \includegraphics[width=0.95\linewidth]{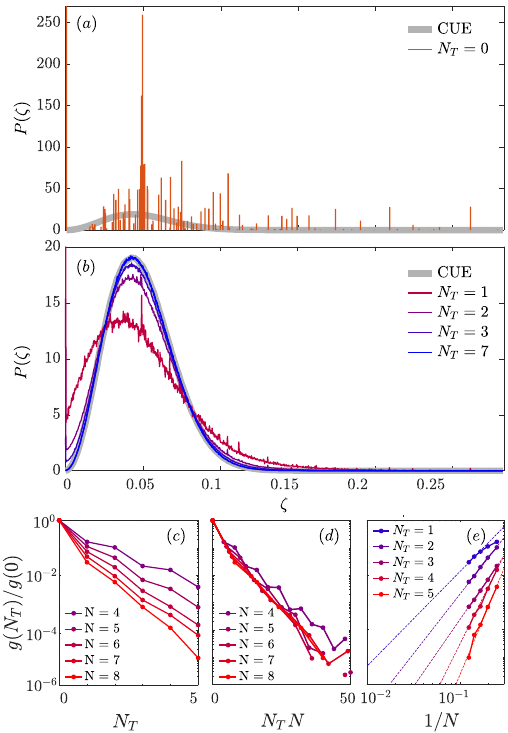}
    \end{center}
    \caption{(a,b) Level spacing distribution of Clifford+$T$ random circuits (color lines) 
    and Haar random unitaries (gray line) for $N=7$ qubits. 
    (a) For pure Clifford circuits ($N_T=0$) the distribution is characterized by a discrete structure 
         and a high eigenvalue degeneracy, corresponding to the feature at $\zeta=0$.
    (b) Injecting $T$-gates,$P(\zeta)$ converges to the CUE distribution.% at $N_T \approx N$. 
     (c-e) Fraction of eigenvalue degeneracy $g(N_T)/g(0)$ for different system sizes $N$ 
     as a function of  $N_T$ (d),  as a function of $N_T N$ (d), and versus $1/N$ for different $N_T\,$s (e). 
    The data suggest that in the thermodynamic limit, $N\to\infty$, any finite number of  $T$-gates 
    eventually resolves all degeneracies (i.e., disrupts all Clifford orbits) for deep-enough circuits.  
    For instance, for $N_T=2$, degeneracies are predicted to be suppressed a millionfold 
    for  $N < 100$ qubits (see dashed line). }
    \label{fig:level_spacing}
\end{figure}
In Figs.~\ref{fig:level_spacing}(a,b) we compare the level spacing distribution distribution $P(\zeta)$ 
for pure Clifford circuits and Haar unitaries. 
For pure Clifford circuits ($N_T=0$), the distribution is characterized by a discrete structure 
and high degeneracies, yielding a delta peak at $\zeta=0$.
Injecting $T$-gates, the degeneracies are resolved, and the distribution converges to the CUE result.
  
Let us now focus on the feature at $\zeta=0$, which corresponds to the fraction of degeneracies $g(N_T)/g(0)$
(see Fig.~\ref{fig:level_spacing}~(c)). In contrast to the number of periodic orbits, the number of degeneracies 
is suppressed \emph{faster} with increasing $N$, and 
our data are suggestive of an exponential scaling with $N\times N_T$.
This is apparently related to the specific structure of periodic orbits. \
The largest periodic orbit is of size $L_{\text{max}}\sim 2^N$, which is exponentially smaller than the number of strings, 
$N_{\text{strings}} = 2^{2N}$. Many of the orbits in undoped Clifford circuits are therefore exponentially large. 
One can argue that these exponentially large orbits contribute an exponentially large fraction to the degeneracy $g(0)$, 
and, moreover, are exponentially sensitive to the presence of $T$-gates, which can destroy them. 
As a result, the number of degeneracies decreases much faster than the number of periodic orbits, and  
is suppressed as $\sim e^{- c N\cdot N_T}$, while the number of orbits 
decreases only as $ \sim e^{- c^\prime N_T}$, with $c$ and $c^\prime$ constants of the order of unity. 
For any finite value of $N_T$,  the fraction of degeneracies is 
suppressed as $N\to\infty$. Indeed, the finite size scaling in Fig.~\ref{fig:level_spacing}~(e)
suggests that in the limit, $N\to\infty$, any finite number of  $T$-gates eventually resolves all but a vanishing fraction of 
degeneracies in the spectrum.
 
We should add that a periodic orbit $P$ in the space of strings, as generated by $\mathbb U$, 
corresponds to a series of eigenphases $\vartheta= \vartheta_P + 2\pi \,m/L$ in the spectrum of $U$,
with $\vartheta_P\in[0,2\pi/L]$ a phase associated with the orbit. 
To have degeneracies in the spectrum of $U$, one therefore needs the phases $\vartheta_P$ of \emph{different} 
orbits to be correlated. Clearly, in the undoped Clifford circuit this is the case.
These observations are also in line with recent results from Haferkampt \textit{et al.}~\cite{Haferkamp2022}, 
which showed that a {\textit{system-size independent}} number $\mathcal{O}\left(k^4\log_2^2(k)\log_2(1/\epsilon)\right)$ 
of non-Clifford gates is sufficient to approximate a $k$-design with precision $\epsilon$. 

%G this is not true, I think
%Note that this is equivalent to the observation of a single  $T$-gate being sufficient 
%to eventually disrupt all periodic orbits for deep-enough Clifford+$T$ circuits. 
%So the two different measures, i.e., the correlation function and the level-spacing statistics, 
%provide the same picture and yield the same conclusion. 

\section{Magic Generation} \label{sec:magic_generation}

%Quantum resource theory~\cite{Chitambar2019}, as a framework for the characterization of quantum states, 
%has recently emerged as a fundamental paradigm within the quantum information community. 
%The underlying idea is that a \textit{resource} is the discriminant for partitioning of all possible quantum states into two groups, 
%free states and resource states. 
%The operations that leaves the free set invariant are referred to as free operations.  
%In quantum physics, the paradigmatic example of quantum resource is entanglement, 
%but also other concepts such as nonlocality, quantum coherence, and quantum correlations 
%have been investigated within the framework of resource theory. 
%
%In particular, a valuable resource in quantum information is \textit{non-stabilizerness} (or \textit{magic}) 
%which represents the hardness to classically simulate a quantum state~\cite{Huang2024}. 
%In the case of magic, the set of stabilizer states (STAB) correspond to the free states, and the Clifford group to the free operations.  
%In the following we focus on the magic generation properties of Clifford+$T$ circuits. 
%Specifically, we characterize the rate of magic growth and the number of  $T$-gates injected in the circuit 
%and the distribution of magic over the ensemble. 

\subsection{Stabilizer R\'enyi Entropy (SRE) and non-stabilizing power} \label{sec:SRE}

%In the case of magic, the set of stabilizer states (STAB) correspond to the free states, and the Clifford group to the free operations. 

%Quantum resource theory~\cite{Chitambar2019}, as a framework for the characterization of quantum states, 
%has recently emerged as a fundamental paradigm within the quantum information community. 
%The underlying idea is that a \textit{resource} is the discriminant for partitioning of all possible quantum states into two groups, 
%free states and resource states. 
%The operations that leaves the free set invariant are referred to as free operations.  
%In quantum physics, the paradigmatic example of quantum resource is entanglement, 
%but also other concepts such as nonlocality, quantum coherence, and quantum correlations 
%have been investigated within the framework of resource theory. 

Within quantum resource theory~\cite{Chitambar2019}, several tools have been proposed 
to classify quantum resources encoded in a quantum state, as a measure of complexity. 
Concepts such as entanglement, nonlocality, quantum coherence, quantum correlations, 
have been investigated within this framework. 
%In this context, the \emph{stabilizer R\'enyi entropy (SRE)}, 
%also referred to as magic, 
%represents the hardness to classically simulate a quantum state~\cite{Leone2022,Huang2024},  
%and has raised significant attention recently~\cite{Oliviero2022,Haug2023,Rattacaso2023,Haug2024,Leone2024,Turkeshi2024,Turkeshi2407.03929,Ahmadi2024}.
In this context, non-stabilizerness identifies  the amount non-Clifford resources 
required to prepare a quantum state~\cite{veitch2014resource}.
It is associated with the hardness to simulate the quantum state classically~\cite{Leone2022,Huang2024}, 
and represents the fundamental resource to unlock quantum advantage. 
Several measures of non-stabilizerness have been proposed in the literature~\cite{veitch2014resource,Howard2017,Heinrich2019,Wang2019,Wang2020,Beverland2020,Liu2022},  
but while most are computationally intractable, the recently introduced \textit{stabilizer R\'enyi entropy} (SRE)~\cite{Leone2022} 
allows for practical numerical and analytical calculations, and has thus raised significant attention \cite{Oliviero2022,Rattacaso2023,Haug2023,Lami2023,Bejan2024,Haug2024,Tarabunga2024,Niroula2024,Turkeshi2024,gidney2024magic,Fux2410.09001,Turkeshi2407.03929}.
In the following, we  rely on SRE to quantify magic, and therefore use these two terms as synonyms. 

Within the stabilizer formalism, the \textit{stabilizer states} are those reachable from the computational basis states 
(e.g., from $\ket{0}^{\otimes N}$) using Clifford operations, 
and can be efficiently represented on a classical computer~\cite{Gottesman1998, aaronson2004improved}. 
By this definition, stabilizer states (elements of STAB) have zero magic. 
To quantify magic in terms of SRE in a pure state $\ket{\psi}$, one first introduces a string probability 
distribution in the basis of Pauli strings, associated with  $\ket{\psi}$  as
\begin{equation}
\Xi_\psi(S) = \frac 1 d  \bra{\psi} S \ket{\psi}^{2},
\end{equation}
where $d=2^N$ is the dimension of the Hilbert space of $N$ qubits. 
The corresponding stabilizer 2-R\'enyi entropy (henceforth simply SRE or magic) associated with this probability distribution reads
\begin{equation} \label{eq:SRE}
    M_2 (\ket{\psi}) = - \log_2 \sum_{S \in \mathcal{P}_N} \Xi^2_\psi(S) - \log_2 (d).
\end{equation}
The SRE also fulfills the following properties~\cite{Leone2022}:
\begin{itemize}
\item[(i)] \textit{faithfulness}: $M_2 (\ket{\psi}) = 0$ iff $\ket{\psi} \in \textrm{STAB}$, otherwise $M_2 (\ket{\psi}) > 0$;
stabilizer states thus correspond to states that are relatively localized in the space of strings and have the smallest possible entropy;
\item[(ii)] \textit{stability under Clifford operations}, $C \in \mathcal{C}_N: \\ M_2 (C\ket{\psi}) = M_2 (\ket{\psi})$;
\item[(iii)] \textit{additivity}: $M_2(\ket{\psi} \otimes \ket{\phi}) = M_2(\ket{\psi}) + M_2(\ket{\phi})$;
and 
\item[(iv)] it is \textit{upper bounded} by $M_2(\ket{\psi}) \leq \log_2((d+1)/2)$. %~\cite{Leone2022}. 
\end{itemize}
We remark that for $N=1$ and $N=3$ qubits, we found states that saturate 
the upper bound (iv) for the SRE, see Appendix \ref{app:A}.

In the context of quantum resource theory, magic is a property of a state. 
Our goal is, however, to characterize the magic generation properties of an \emph{operator}
or, rather, a class of operators, the ensemble of Clifford+$T$ circuits. 
A step in this direction has been done with the \textit{non-stabilizing power} of a unitary operator $U$~\cite{Leone2022}
 \begin{equation} \label{eq:non-stabilizing-power}
  {\mathcal M}_2(U) = \frac{1}{|\textrm{STAB}|} \sum_{\ket{\psi} \in \textrm{STAB}} M_2\left(U\ket{\psi}\right), 
 \end{equation}
which associates an average value of SRE to any given unitary operator. 
This quantity measures the power of a quantum circuit to generate complex states from stabilizer states.
For the non-stabilizing power, some interesting results are available~\cite{Leone2022}, including: 
(i) a lower bound for the typical non-stabilizing power for Haar unitaries 
$\langle \mathcal{M}_2(U) \rangle \geq \log_2\left((d+3)/4\right)$, and
(ii) a lower bound on the number of  $T$-gates needed in addition to Clifford circuits 
to decompose a unitary operator $U$, which is found to be $\gtrsim O(N)$. 

Computing the non-stabilizing power of a quantum circuit $U$ is very expensive, 
as it requires evaluating $M_2$ as in Eq.~\eqref{eq:SRE}
-- which is already a numerically expensive endeavor -- and then averaging it 
over the exponentially large set of stabilizer states, see Eq.~\eqref{eq:non-stabilizing-power}. 
Furthermore, in the case of random circuits, an averaging over the ensemble  
also needs be performed.

Here we complete this program: we investigate the rate of magic growth in Clifford + $T$ circuits as a function of injected $T$-gates.
We compute numerically the distribution of magic $\rho(M_2)$ for the ensemble, 
and compute the average non-stabilizing power $\langle {\cal M}_2\rangle_U$
as a function of system size, $N$, and the number of $T$-gates, $N_T$.

\subsection{One Qubit} \label{sec:magic_one-qubit}

Before showing the results for more complex Clifford+$T$ circuits, let us focus on the simplest case 
of a single qubit, and study the magic of its states, as well as the non-stabilizing power of 
one-qubit Haar-random gates. While being the simplest instance, analyzing this provides useful tools 
to understand the magic generation properties of more complex circuits.
  
%While being the simplest case, analyzing this case provides useful tools 
%to understand the magic generation properties of more complex circuits, 
%where the magic-generating unit is a non-Clifford single-qubit gate (T in this specific case).
We parametrize  $1$-qubit states on the Bloch sphere as 
\begin{equation} \label{eq:Bloch}
 \ket{\psi} = \cos(\theta/2)\ket{0} + e^{i\varphi} \sin(\theta/2) \ket{1}, 
\end{equation}
and calculate the magic $M_2(\theta, \varphi)$ on a dense mesh of polar and azimuthal angles, 
$0~\leq~\theta~\leq~\pi$, and $0~\leq~\varphi~<~2\pi$, while weighting with the Haar measure, 
$\sim \sin(\theta) \rmd \theta \rmd\phi$. 

\begin{figure}[tp!]
    \centering
    \includegraphics[width=0.95\linewidth]{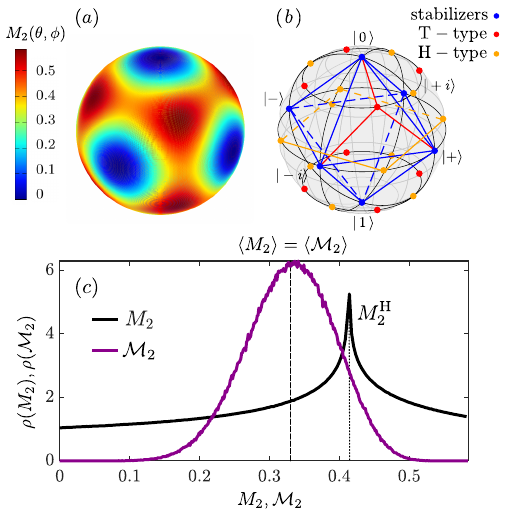}
    \caption{(a) Color plot of magic on the Bloch sphere for a single qubit. 
    (b) Special states on the Bloch sphere: states with magic $M_2(|\psi\rangle)=0$,
     i.e., stabilizer states occupy the vertices of an octahedron (blue),
    $\ket{0}, \ket{1}, \ket{+}, \ket{-}, \ket{+i}, \ket{-i}$;
    $T$-type states $\ket{T}$ possess maximal magic, $M_2 = \log_2 (3) -\log_2 (2) \approx 0.585$ (red); 
    $H$-type states $\ket{H}$ are associated with the most probable value of magic $M_2 = M_2^{\mathrm{H}}$ (orange), 
    see text for a discussion. 
    (c) Distribution of magic, $\rho(M_2)$, obtained by sampling $1$-qubit states uniformly on the Bloch sphere, 
    and distribution of the non-stabilizing power $\rho(\mathcal{M}_2)$ of 1-qubit Haar-random unitaries. 
    The two distributions have the same mean value, $\langle M_2 \rangle_{\psi} = \langle {\cal M}_2\rangle_U$, 
    % $\mathbb{E}(\mu)$ 
    but $\rho(M_2)$ has a logarithmic singularity at $M_2^{\mathrm{H}} \approx 0.414$ corresponding 
    to the $H$-type states, $\ket{H}$. }
    \label{fig:one-qubit_magic}
\end{figure}

Note that this is equivalent to sampling the unitary operators $U \in \mathcal{U}(2)$ with uniform Haar measure, and applying $U$
to the stabilizer state $|0\rangle $, or, equivalently, picking states uniformly on the Bloch sphere.
We show the magic map on the Bloch sphere in Fig.~\ref{fig:one-qubit_magic}(a), 
and the corresponding probability density $\rho(M_2)$ in Fig.~\ref{fig:one-qubit_magic}(c). 
The maximum achievable magic value we find coincides with the theoretical upper bound~\cite{Leone2022},
$M^{\rm max}_2 = \log_2 ((d+1)/2) = \log_2 (3/2) = 0.5849\dots$, 
proving that the bound is tight for $N=1$. 

%We identify a few interesting states on the Bloch sphere, 
It is possible to identify a few interesting states on the Bloch sphere~\cite{Anwar2012}, 
as shown in Fig.~\ref{fig:one-qubit_magic}(b).  
As expected, states with zero magic correspond to the one-qubit stabilizer states 
$\{\ket{0}, \ket{1}, \ket{+}, \ket{-}, \ket{i}, \ket{-i} \}$,
which occupy the vertices of an octahedron.  
Clifford group elements generate symmetry operations on this octahedron~\cite{Virmani2005,Anwar2012}. 
States with maximal SRE, $M^{\rm max}_2 = M_2^{\mathrm{T}} = M_2(\ket{T}) = 0.585\dots$   
are identified as $T$-type states
(red dots in Fig.~\ref{fig:one-qubit_magic}(b))~\cite{Cepollaro2024}. 
%which correspond to the radial projection of the center of the faces of the octahedron on the Bloch sphere.  
A representative $T$-type state can be generated as $\ket{T}=THTH\ket{0}$, where $H$ denotes the Hadamard gate. 
The other $T$-type states (8 in total) can be obtained from this by applying phase gates $S$ ($\pi/2$ phase gate) %rotation), 
or by combining Pauli $Y$  and  $S$ gates. %, in both cases a four-fold rotation wraps around the sphere. 
Another set of interesting states is that of $H$-type states (orange dots in Fig.~\ref{fig:one-qubit_magic}(b)).
%denoted as $\ket{H}$. 
On the magic map in Fig.~\ref{fig:one-qubit_magic}~(a), these correspond to saddle points
with a magic $M_2^{\mathrm{H}} = M_2(\ket{H}) = 0.414\ldots$. 
This is identified as the most likely magic,
associated with  the van Hove singularity of the underlying saddle point.
A representative $H$-state on the equator can be generated as $\ket{H} = TH\ket{0}$, and 
all other $H$-states (12 in total) can be generated by applying on this state Clifford gates $H$, $Y$ and $S$.
%
%whereas (ii)  $HTH\ket{0}$ and (iii) $YHTH\ket{0}$ generates the states in the two hemispheres. 
%All other H-type states can be obtained from those by applying (an appropriate number of times) the $S$ gate. 
%The SRE of these states correspond to the most probable value 
%$M_2^{\mathrm{H}} \approx 0.414$ for the distribution $\rho(M_2)$. 
Interestingly, %it has been shown that 
these states can be obtained 
with a protocol known as \textit{magic state distillation}, 
and allow for universal quantum computing~\cite{Bravyi2005,Anwar2012,krishna2019towards,bao2022magic}.

Finally in Fig.~\ref{fig:one-qubit_magic}(c) we show the one-qubit magic
distribution $\rho(M_2)$. 
The distribution is nearly featureless, except for a logarithmic singularity at the most-probable value $M_2^{\mathrm{H}}$, 
corresponding to the magic of $H$-type states. 
This can be understood as a van Hove singularity due to the saddle point.
%generated by applying a $T$-gate  to any of the eigenstates of Pauli $X$ and $Y$
%~\footnote{Analogous result are obtained applying a $\pi/4$ rotation 
%around the $X$ ($Y$) axis to the eigenstates of Pauli operator lying on the corresponding equator, i..e, $Y$ and $Z$ ($X$ and $Z$).}.
For comparison, we also show the distribution of the non-stabilizing power $\rho(\mathcal{M}_2)$
of Haar random single-qubit operators $U \in \mathcal{U}(2)$.  This we obtained by sampling $U$,
and calculating the corresponding non-stabilizing power ${\mathcal M}_2(U)$ from its definition, Eq.~(\ref{eq:non-stabilizing-power}). 
%As mentioned, this calculation is very expensive for $N$ qubits. 
The distributions of $M_2$ and $\mathcal{M}_2$ display quite different shapes, 
as $\rho(\mathcal{M}_2)$ lacks the characteristic singularity observed in $\rho(M_2)$, 
which is washed away by averaging over the stabilizer states. 
The two distributions, 
however,  have the \textit{same} expectation value, $\langle M_2 \rangle_\psi = \langle {\cal M}_2\rangle_U$.
%denoted by $\mathbb{E}(\mu)$ in  Fig.~\ref{fig:one-qubit_magic}(c).  

\subsection{Magic generation in Clifford+$T$ circuits} \label{sec:magic_Clifford+$T$}

We investigate  magic generation in Clifford+$T$ circuits~\cite{li2023optimality,Bejan2024} by applying the following protocol: 
\begin{itemize}
\item[(i)] We initialize the qubit register in a random stabilizer state $\ket{\psi} = C (\ket{0}^{\otimes N})$,  
where $C$ is a brick-wall Clifford circuit deep enough to represent an element of $\mathcal{C}_N$. 
By construction, this initial state has no magic, $M_2(\ket{\psi})=0$. 
\item[(ii)] Generate a random Clifford+$T$ circuit $U$ with a given depth $D$ and number of injected  $T$-gates, $N_T$.
 \item[(iii)]  We calculate the magic of this state, $M_2(U\ket{\psi})$.
\item[(iv)]  We obtain the average non-stabilizing power of the ensemble, $\langle {\cal M}_2\rangle_U$
 by averaging $M_2(U\ket{\psi})$ over  $N_s=2^{16}$  initial states and random circuits. 
% calculate the magic (SRE) of the state as $M_2(U\ket{\psi})$, 
 %The protocol is repeated for $N_s=2^{16}$ samples to extract the ensemble properties. 
\end{itemize}
Notice that the procedure above performs a simultaneous average over the stabilizer states as well as 
the circuit ensemble. Therefore, assuming self-averaging, 
the average magic of the final states is equal to the average stabilizing power 
of the circuits, 
\begin{equation}
 \langle \,{M}_2(U|\psi\rangle) \,\rangle_{U,\psi\in \rm STAB} = \langle {\cal M}_2\rangle_U\;.
\end{equation}
This has been explicitly demonstrated in the previous subsection for $N=1$. 
We remark that this protocol is slightly different from the one in Ref~\cite{haug2024Probing}; 
there, completely random Clifford-circuits have been assumed between two $T$-gates. 
In contrast, here we use a sufficiently deep brick-wall structure, which we dope 
eventually with a relatively large number of $T$-gates. 
Therefore, in the regime where the number of $T$-gates is larger than the depth of the circuit, $N_T > D$, 
the number of layers separating two $T$-gates is typically small compared to the number of qubits, 
and the Clifford operator corresponding to this layer is still not entirely random.

Let us note that the numerical evaluation of the magic is exponentially expensive in the system size, 
since the number of (unsigned) Pauli strings $P \in \mathcal{P}_N$ scales as $|\mathcal{P}_N|=4^N$. 
It has been shown that the SRE can be computed efficiently for certain classes of states
such a matrix product states~\cite{Haug2023,Lami2023,Tarabunga2024}, 
but not for the general states discussed here. 
This poses severe limitations to the system sizes investigated within our framework.

\begin{figure}[t!]
    \centering
    \includegraphics[width=0.95\linewidth]{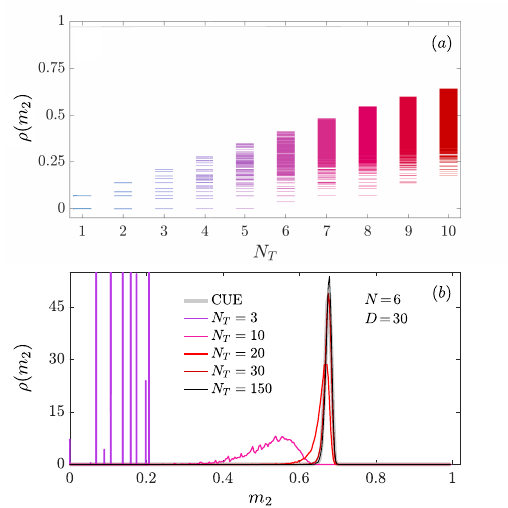}
    \caption{Evolution of the magic density distribution in Clifford+$T$ circuits for $N=6$ qubits and $D=5N$. 
    (a) Spectrum of the normalized SRE as a function of $N_T$ (lines from blue to red). 
    For a few injected  $T$-gates the spectrum is discrete, but eventually becomes quasi-continuous as $N_T$ increases. 
    (b) Magic density distribution %$\rho(\mu)$ 
    $\rho(m_2)$ for selected values of $N_T$.
    The limiting distribution, i.e., for fully-saturated circuits with $N_T \to D \,N$,  
    converges to the numerically sampled distribution for Haar unitaries (CUE). 
    For $N=6$, most of the weight is found in a sharp peak at a value that is well below the theoretical maximum. 
   }
    \label{fig:magic_distribution}
\end{figure}

Fig.~\ref{fig:magic_distribution} shows the evolution of the probability distribution 
of the normalized or \emph{magic density}, $m_2 \equiv  {M}_2 / N$. %$\mu \equiv  {M}_2 / N$. 
We have sampled the probability density %$\rho(\mu)$ 
$\rho(m_2)$  numerically for different numbers of randomly positioned  $T$-gates, $N_T$, 
while acting on stabilizer states with Clifford+$T$ circuits. 
In the case of a single $T$-gate, $N_T=1$, we find that the distribution is \textit{discrete} and \textit{bimodal}, 
and the only two possible normalized SRE values are $m_2=0$ %$\mu=0$  
and, %$\mu_{H}$} 
$m_2^{\mathrm{H}}=M_2^{\mathrm{H}}/N$. 
The \textit{magic quantum} $M_2^{\mathrm{H}}$ corresponds to the SRE of a single qubit H-type state, 
%or, more in general, to 
also identified as the magic generated 
by a single $T$-gate acting (non-trivially) on a stabilizer state~\cite{haug2024Probing} 
%The magic jumps we observe are indeed close to the value $???$, predicted in Ref~\cite{haug2024Probing}.
As $N_T$ increases, the spectrum of possible magic values increases exponentially, 
but it non-trivially preserves its discrete character for small $N_T$. 
Eventually, for an \textit{extensive} number of injected  $T$-gates, the magic displays a quasi-continuous spectrum, 
and low-magic values become statistically unlikely (see Fig.~\ref{fig:magic_distribution}(a)). 
The corresponding distribution %$\rho(\mu)$ 
$\rho(m_2)$ becomes skewed and peaked at higher values, 
which nevertheless remain significantly below the theoretical upper bound for $N=6$. 
%For the brick-wall Clifford+$T$ circuits with a finite depth, there exists a \textit{fully-saturated} limit, 
%corresponding to $N_T=N D$ injected  $T$-gates.  
Finally, in the limit $N_T \gg N$, we find that %$\rho(\mu)$ 
$\rho(m_2)$ approaches the magic distribution of states generated by Haar random unitaries, with an accuracy %threshold 
limited by statistical noise due to sampling.  
The magic distribution is qualitatively similar to the one in Fig.~\ref{fig:magic_distribution} for all $N\ge 5 $ values studied here.

\begin{figure}[t!]
    \centering
    \includegraphics[width=0.95\columnwidth]{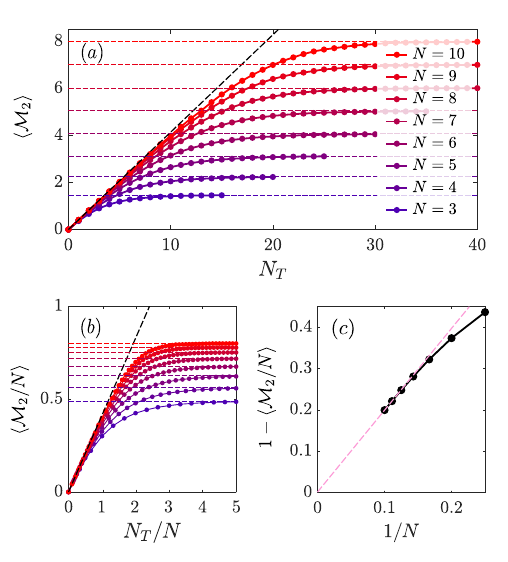}
    \caption{Magic generation of Clifford+$T$ circuits for different system sizes $N$, and depth $D=5N$.  
    (a) In the sparse limit, $N_T \lesssim N$, the average non-stabilizing power scales as 
    $\langle {\cal M}_2  \rangle = \langle M_2 \rangle \approx {\cal M}^\infty_2 (T) N_T$,  
    with $\mathcal{M}^\infty_2(T) \approx 0.414 $ the average non-stabilizing power 
    of a single $T$-gate acting on a  system of $N\to\infty$ quibits.
    Horizontal dashed lines denote the average non-stabilizing power computed for Haar random unitaries.
    (b) The average non-stabilizing power density $ \langle {\cal M}_2 / N \rangle $ as a function of $N_T / N$. 
    In the limit of $N_T \lesssim N$ displays a universal slope. 
    In the limit $N_T \gg N$, it recovers the value calculated for Haar-random unitaries (dashed lines). 
    (c) Asymptotic values of  $\langle {\cal M}_2 /N\rangle$, as computed for Haar-random unitaries, 
    plotted against $1/N$. }
    \label{fig:magic_generation}
\end{figure}

%The behavior of the non-stabilizing power of the Clifford+$T$ circuits agrees with the findings of Ref.~\cite{haug2024Probing}. 
To close this section, let us discuss the behavior of the average non-stabilizing power of  Clifford+$T$ circuits.
As shown in Fig.~\ref{fig:magic_generation}(a), the average magic 
%generated by  Clifford+$T$ circuits 
scales approximately linearly with the number of injected  $T$-gates in the regime $N_T \lesssim N$, 
i.e., $\langle {\cal M}_2\rangle \approx N_T  \,{\cal M}^{\infty}_2(T)$,  
with ${\cal M}^{\infty}_2(T) = 0.414 \ldots$,  
the non-stabilizing power of a single $T$-gate applied on a large system of qubits, $N\to\infty$.  
%Astonishingly, within our numerical accuracy, the value ${\cal M}^{\infty}_2(T)$ seems to coincide with the 
%magic of a single quibit $|H\rangle$-state, ${\cal M}^\infty_2(T) = M_2^{N=1}(|H\rangle)$. 
%$M_2^{\mathrm{H}}$, 
%the most probable value of the single-qubit magic distribution $\rho(M_2)$, see Fig.~\ref{fig:one-qubit_magic}(c). 
%The reason behind this is that, one the one hand, in large circuits, 
%$T$-gates leave only a small fraction of stabilizer states invariant, 
%and transform most of them into $H$-like states, 
%and on the other hand, remote $T$ gates act independently. 
%Therefore, in the sparse $T$-gate limit ($N_T \lesssim N$) %, $T$-gates do not interfere, and 
%magic is generated in approximate quanta of size $M_2^{N=1}(|H\rangle)$.
%
Magic, however, cannot increase indefinitely:  for $N_T \gtrsim N$
"interference" between multiple $T$-gates slows down the magic generation rate, 
and the average magic, i.e., the average non-stabilizing power of the random circuits, 
$\langle {\cal M}_2\rangle$, saturates
%It is natural to expect that the saturation value corresponds 
at the average non-stabilizing power of Haar-random unitary circuits, 
%This is confirmed by numerical simulations where we averaged over Haar-random unitary circuits, 
(see dashed horizontal lines in Figs.~\ref{fig:magic_generation}(a,b)). 
This suggests that the typical complexity of a Haar-random circuit can be reached by $N_T \sim  {\cal O}(N)$ $T$-gates. 
Our numerical simulations reproduce the analytical result of Ref.~\cite{haug2024Probing} 
in both the $N_T \ll N$ and $N_T \gg N$ limits apart from minor deviations in the 
crossover regime, $N_T \sim N$, due to the relatively shallow Clifford layers separating the $T$-gates, 
as discussed above. 
%
%Clearly, for $N_T\gg N$, the average non-stabilizing power $ {\langle {\cal M}_2 \rangle}$ saturates at a value corresponding to 
%the average non-stabilizing power of Haar random circuits, indicating 
%
Fig.~\ref{fig:magic_generation}(b) displays the non-stabilizing power \emph{density}, 
%$m_2 \equiv \mathcal{M}_2(U)/N$,  
as a function of  $T$-gate density, $N_T/N$. 
In agreement with Ref.~\cite{haug2024Probing}, we find that the average magic density 
approaches that of Haar-random unitaries. %circuits.  
The latter is bounded from below as 
$\langle {m}_2(U)\rangle_U  \geq \frac 1 N  \log_2((2^N+3)/ 4)$~\cite{Leone2022}, 
%In particular, $\langle \mu \rangle \geq \log_2((d+3)/4)/N = 0.677\dots$ for $N=6$. 
which approaches unity as $1 - {2}/{N} + O(2^{-N})$, in the thermodynamic limit, $N \to \infty$. 
This relatively large finite size correction is demonstrated in Fig.~\ref{fig:magic_generation}~(c), 
where we plot %$1-\langle \mu \rangle$ 
$1 - \langle m_2 \rangle$, 
as a function of $1/N$, as computed for Haar-random circuits. 
%Our data are thus consistent with a slow, $\sim 1/N$  convergence,  $\langle \mu\rangle\to 1$.
Our numerical data as well as simple analytical arguments support the emergence of a phase transition~\cite{haug2024Probing}: 
above a critical concentration, $n_T^\star\approx 2.41$, the non-stabilizing power of a 
deep random $T$-doped Clifford circuit becomes unity in the thermodynamic limit, 
while below it magic is simply proportional to $n_T$,  and $ \langle {m}_2\rangle_U = n_T/ n_T^\ast$.
%As stated in Ref.~\cite{haug2024Probing}, circuit construction protocol.

\section{Conclusions and Outlook} \label{sec:discussion}

In this work, we examined how complexity is reflected in the 
 spectral properties and magic generation capabilities
of random $T$-gate-doped Clifford circuits.
For pure Clifford circuits, $C$, we uncovered a unique periodic orbit structure in the space of
Pauli strings, which gives rise to large degeneracies in the spectrum as well as delta-correlations. 
These  properties reflect the inherent simplicity of Clifford circuits, which, despite their capacity to generate 
entanglement, remain efficiently simulable due to their lack of non-stabilizerness or "magic." 
At the same time, these spectral features turn out to be sensitive indicators of (non)-complexity; 
they capture accurately how periodic orbits and corresponding spectral degeneracies are destroyed 
upon $T$-doping.

For a finite number of $T$-gates, $N_T$, and a finite number of qubits, $N$, spectral correlations as well 
as level spacing statistics reveal the simultaneous development of chaotic features described by random-matrix theory, 
coexisting with exponentially suppressed Clifford-anomalies. 
For the deep circuits studied here, $D\gg N$, the number of periodic orbits is reduced 
with the number of $T$-gates as $e^{- c\,N_T}$, while the degeneracy anomaly in the level spacing distribution, 
is suppressed roughly as $e^{- c'\,N_T N}$. Therefore, in the thermodynamic limit, even a single 
$T$-gate is sufficient to drive the entire spectrum of the circuit towards chaotic 
behavior, as captured by RMT.

We compared this behavior with the emergence of complexity in terms of magic generation, 
%– referred to as stabilizer Renyi entropy (SRE)~\cite{Leone2022}. 
quantified in terms of stabilizer Renyi entropy (SRE)~\cite{Leone2022}. 
The corresponding quantity, the \emph{non-stabilizing power}, which we could refer to as the 
\emph{magicness of the circuit}, characterizes the complexity of the quantum circuit. 
To this end, we obtained the distribution of magic density $\rho(m_2)$ that characterizes the random circuit ensemble. 
In the case of Clifford+$T$ circuits, we find that for a single injected $T$-gate, 
the distribution is discrete and bimodal, 
with the two possible values being $m_2=0$ and $m_2^{\mathrm{H}}=M_2^{\mathrm{H}}/N$. 
For $N_T \ll N$, the distribution partially preserves its discrete character. 
%as the values of magic cluster around (approximate) integer multiples of $M_2^{\mathrm{H}}$. 
Increasing $N_T$ further, the distribution becomes quasi-continuous, 
but it is skewed and strongly peaked towards higher values of SRE. 
Eventually, in the limit of $N_T \gg N$ we find that $\rho(m_2)$ converges to the distribution of the Haar unitary ensemble, 
and displays a exponentially peaked~\cite{szombathy2025} distribution concentrated around ${m}_2 \approx 1-2/N$. 

Averaging the distribution, we can access the average non-stabilizing power 
of the circuit ensemble, $\langle \mathcal{M}_2 \rangle$. 
Classically simulable Clifford circuits have vanishing non-stabilizing power, 
and magic is generated by injecting $T$-gates into them. 
%
%In particular, classically simulatable Clifford circuits have vanishing non-stabilizing power.
%
%Here we investigated, how introducing $T$-gates to Clifford circuits impacts 
%the distribution of magic in the final states, and the (average) non-stabilizing power of the circuits.  
%
In the dilute limit, $N_T \ll N$, the interplay between $T$-gates is statistically irrelevant, 
and the magic generation properties can be entirely understood 
in terms of the magic generated by a single $T$-gate on typical stabilizer states, 
$\mathcal{M}^\infty_2(T) = 0.414\dots$~\cite{haug2024Probing}. 
Our numerics confirms that the average magic – and thus the average non-stabilizing power of a circuit – 
increase as $\langle {\cal M}_2\rangle \approx {\cal M}^\infty_2(T)\, N_T$ for $N_T\ll N$, 
%The average magic and the average non-stabilizing power of a circuit 
%$\langle {\cal M}_2\rangle$ 
and saturate at $N_T\gtrsim N$, at the non-stabilizing power of Haar-random unitaries, 
$\langle {\cal M}_2\rangle_\text{Haar} \approx N-2$~\cite{haug2024Probing,szombathy2025}. 

Our numerics are thus consistent with recent analytical results~\cite{haug2024Probing}
that $\langle m_2 \rangle_{n_T>n_T^*}  = 1$ in the thermodynamic limit 
above a critical $T$-gate concentration, $n_T^*\approx 2.41 $, 
while it is simply linear below this, $\langle m_2\rangle_{n_T<n_T^*} =  n_T/n_T^*$. 
The critical concentration $ n_T^*$ is simply related to the magic generated by a single $T$-gate, 
$n_T^*=1/{\cal M}^\infty_2(T)$.

These findings highlight the interplay between simplicity and complexity in quantum circuits, providing 
an understanding of how adding non-Clifford elements induces chaotic and resource-rich behaviors.

%!They indicate further that generic unitary operations may be approximated by Clifford+$T$ circuits with 
%an extensive but still limited number $N_T\sim N$ of $T$-gates. 

\acknowledgments
We thank A.~Hamma for insightful discussions. 
This research was supported by the Ministry of Culture and Innovation and the National Research, Development and Innovation Office (NKFIH) within the Quantum Information National Laboratory of Hungary
(Grant No. 2022-2.1.1-NL-2022-00004), through NKFIH research grants Nos. K134983, SNN139581, 
and QuantERA `QuSiED' grant No. 101017733.
D.S. acknowledges the professional support of the doctoral student scholarship program of the co-operative doctoral program of the Ministry for Innovation and Technology from the source of the National Research, Development and Innovation fund.
C.P.M. acknowledges support by the Ministry of Research, Innovation and Digitization, CNCS/CCCDI–UEFISCDI, 
under the project PN-IV-P1-PCE-2023-0159. 
T.R. was supported by the HUN-REN Welcome Home and Foreign Researcher Recruitment Programme 2023. 
We acknowledge KIFÜ 
%(Governmental Agency for IT Development, Hungary, \url{https://ror.org/01s0v4q65}) 
for awarding us access to the Komondor HPC facility based in Hungary.

\appendix
\section{Maximum magic states}\label{app:A}

The theoretical upper bound of the SRE was determined by maximizing the sum corresponding to the projection of state $\ket{\psi}$ 
on the $N$-qubit Pauli basis, with weights $\Xi_{I} = d^{-1}$ and $\Xi_{P \in \mathcal{P}_N/I} = d^{-1} (d+1)^{-1}$~\cite{Leone2022}. 
%i.e., $\ket{\psi}\bra{\psi} = d^{-1} \sum_P \bra{\psi} P \ket{\psi} P$~\cite{Leone2022}.
We identified states that satisfy these criteria for $N = 1$ and $N = 3$ qubits, 
while demonstrating that these criteria cannot be satisfied for $N = 2$ qubits.

For a single qubit, the maximum magic state can be found by writing down the equations for the SRE 
for a generic $\Xi_{P}$ and solve it for the criterion above. 
From symmetry considerations, it immediately follows that the state with maximum magic $M_2=\log_2(3/2)$ 
%takes the form
is defined in terms of the projector~\cite{bravyi2005universal}
\begin{equation}
    %\ket{\psi} = \frac{1}{\sqrt{2}} \left( \ket{0} + e^{i\pi/4}\ket{1} \right) = \ket{T},
   \ketbra{\psi} = \frac{1}{2}\left(\Id+\frac{X+Y+Z}{\sqrt{3}}\right) = \ketbra{T}, 
\end{equation}
or equivalently, with the parametrization of Eq.~\eqref{eq:Bloch} 
\begin{equation}
   \ket{\psi} = \cos(\theta^*/2)\ket{0} + e^{-i\pi/4} \sin(\theta^*/2) \ket{1} = \ket{T}, 
\end{equation}
where $\theta^*=\arccos(1/\sqrt{3})$, 
and it corresponds to one of the 8 T-type states the Bloch sphere, as shown in Fig.~\ref{fig:one-qubit_magic}(b). 

For $N = 2$, our analysis of the system of equations yields contradicting constraints. 
Hence, a state that possess the theoretical upper bound magic $M_2=\log_2(5/2)$ does \textit{not} exist.

For $N = 3$ qubits we have employed a combinations of Monte Carlo and variational approaches. 
These led us to a set of states of simple structure 
\begin{equation}
    \ket{\psi} = \mathcal{N} \sum_j c_j \ket{j},
\end{equation}
where $c_j \in \left\lbrace 1, i, 1+i, 0 \right\rbrace$, and $\mathcal{N}$ is the appropriate normalization. 
We have identified 448 such states for $N = 3$ which possess magic $M_2=\log_2(9/2)$. 
A representative state of this set takes the form
\begin{equation}
    \ket{\psi} = \frac{1}{\sqrt{6}} \Big( \ket{000} + \ket{001} + \ket{010} + i\ket{011} + (1 + i)\ket{100} \Big).
\end{equation}

\bibliography{references}

\end{document}